\documentclass[reprint,amsmath,amssymb,aps,prl,superscriptaddress,nofootinbib,floatfix]{revtex4-2}
\usepackage{graphicx}
\usepackage{dcolumn}
\usepackage{bm}
\usepackage{mathrsfs}
\usepackage{amsmath}
\usepackage{hyperref}
\usepackage{latexsym}
\usepackage{array}
\usepackage{multirow}
\usepackage{booktabs}
\usepackage{tikz}
\usetikzlibrary{patterns,decorations.pathmorphing,shapes}
\hypersetup{colorlinks=true, linkcolor=blue, urlcolor=blue, citecolor=magenta}

\begin{document}

\title{Anisotropic Electrostatic-Elastic Softening and Stability in Charged Colloidal Crystals}
\author{Hao Wu}
\email{wuhao@ucas.ac.cn}
\affiliation{Zhejiang Key Laboratory of Soft Matter Biomedical Materials, Wenzhou Institute, 
University of Chinese Academy of Sciences, Wenzhou, Zhejiang 325000, China}
\author{Zhong-Can Ou-Yang}
\affiliation{Institute of Theoretical Physics, Chinese Academy of Sciences, Beijing 100190, China}

\date{\today}

\begin{abstract}
Charged colloidal crystals exhibit a subtle interplay between electrostatic screening and elastic deformation. 
{When an isotropic electrostatic volumetric softening acts on an anisotropic elastic background, the longitudinal acoustic response softens preferentially along specific crystallographic axes.} 
This article provides a self-contained derivation of the long-wavelength {acoustic stability} condition for cubic crystals subject to a generic electrostatic-elastic coupling. 
Starting from an effective static elastic tensor renormalized by a scalar coupling constant $\lambda_g$, we obtain an explicit condition for the onset of a {long-wavelength acoustic} instability: the direction $\hat{\mathbf{k}}$ that first loses {strong ellipticity} is determined by the inverse Christoffel matrix evaluated along that direction. 
Closed-form expressions for the critical coupling $\lambda_g^c$ are given for the $[100]$, $[110]$, and $[111]$ high-symmetry directions. 
We further provide a microscopic derivation of $\lambda_g$ from the Poisson-Boltzmann theory in a spherical Wigner-Seitz cell, linking the phenomenological constant to experimentally accessible parameters such as salt concentration, particle charge, and volume fraction. 
The analysis reveals that the most fragile direction can be identified without full lattice-dynamical calculations, and the associated unstable strain patterns are discussed. 
{We also compare these acoustic thresholds with the Born stability condition for homogeneous volume change and show that, for a Born-stable reference cubic crystal, the volume mode becomes unstable before any directional acoustic mode, provided homogeneous dilation is admissible.} 
{Numerical illustrations using representative modulus values reported for soft colloidal assemblies illustrate how the criterion can be applied to predict directional softening trends.} 
The present framework serves as a diagnostic tool for interpreting directional anomalies in static compressibility or low-frequency acoustic softening.
\end{abstract}

\maketitle

\section{Introduction}

Colloidal crystals assembled from charged particles are a remarkable platform for studying elasticity, electrostatics, and their mutual feedback \cite{Russell2015, Sutton2020, Pieranski1983, Lindsay1982, Alexander1984, Sood1991, Russell1989}. 
In such systems the Debye screening length, which governs the range of electrostatic interactions, is comparable to the lattice spacing{~\cite{Lindsay1982, Alexander1984}}. 
When the crystal is deformed, local volume changes alter the confining volume available to mobile counterions, thereby modifying the local screening environment. 
This electrostatic-elastic coupling renormalizes the effective elastic moduli and can even drive a static mechanical instability \cite{Wu2025,Wu2024EPL,Wu2026electrostatic,Robbins1988, Leunissen2005, Yethiraj2003}. 
The fundamental mechanisms of structural ordering and crystallization in such soft matter systems are well documented \cite{Gasser2001, Aastuen1986}.

While the isotropic version of this coupling has been studied in detail \cite{Wu2025, Wu2024EPL}, real colloidal crystals are rarely isotropic. 
Face-centered cubic (fcc) and body-centered cubic (bcc) structures, as well as engineered superlattices of DNA-coated nanoparticles \cite{Park2008, Macfarlane2011}, 
possess elastic anisotropy characterized by three independent moduli $C_{11}$, $C_{12}$, and $C_{44}$. 
Consequently, the softening induced by electrostatic-elastic feedback is direction dependent: the crystal may become unstable along a particular crystallographic axis before others. 
Identifying the most fragile direction is therefore essential for interpreting experiments such as Brillouin spectroscopy or static strain-response measurements.

{
A full continuum theory of charge-elasticity coupling has been established for isotropic media in Refs.~\cite{Wu2025,Wu2024EPL}, and the nonlinear Poisson-Boltzmann (PB) cell model was treated by Alexander \emph{et al.}~\cite{Alexander1984}, who showed that charge renormalization and nonlinear screening can lead to a softening of the effective bulk modulus, i.e.\ a positive $\lambda_g$. 
In the present work we extend that framework to cubic anisotropy and focus on extracting practically useful analytical criteria for the onset of a long-wavelength instability. 
We do not attempt to solve the full nonlinear electrostatic-elastic problem; instead we introduce a phenomenological scalar coupling $\lambda_g$ and derive its critical value from the bare elastic constants. 
To make the connection with measurable quantities, we provide a microscopic derivation of $\lambda_g$ within the Debye-H\"uckel (DH) limit of the PB cell model, and we discuss the conditions under which $\lambda_g>0$ (softening) is realized. 
The novel contributions of the present work are threefold: (i) We derive closed-form expressions for the critical electrostatic-elastic coupling $\lambda_g^c$ along the three principal high-symmetry directions of a cubic crystal, Eqs.~\eqref{eq:lam3}. (ii) We provide a microscopic derivation of the coupling constant $\lambda_g$ from the PB theory in a spherical cell model, Eq.~\eqref{eq:lambda_def}, thereby linking the phenomenological parameter to measurable quantities such as salt concentration and particle charge. (iii) We construct a phase diagram in the space of reduced elastic constants, Fig.~\ref{fig:phasediagram}, which identifies the softest direction for arbitrary cubic moduli and discuss the corresponding unstable strain patterns. The resulting formulas involve only the three cubic elastic constants and the coupling strength $\lambda_g$. 
They provide a straightforward diagnostic tool that can be applied once the elastic moduli are known, without requiring a full solution of the coupled PB-elasticity equations.

It is important to clarify from the outset the physical meaning of the elastic constants $C_{ijkl}$ appearing in our model and their relation to the electrostatic-elastic coupling $\lambda_g$. 
We decompose the total free energy of the charged colloidal crystal as
\begin{equation}
\mathcal{F}_{\mathrm{tot}} = \mathcal{F}_{\mathrm{ref}} + \mathcal{F}_{\mathrm{es}},
\end{equation}
where $\mathcal{F}_{\mathrm{ref}}$ is a reference free energy that includes all contributions not explicitly contained in the PB cell calculation. The remaining part, $\mathcal{F}_{\mathrm{es}}$, is the electrostatic free energy computed from the nonlinear PB cell model. It consists of the electrostatic self-energy of the complete charge distribution (mobile ions plus fixed surface charge) and the mixing entropy of the mobile ions, as detailed in Sec.~II. The bare elastic constants $C_{ijkl}$ are defined from this reference free energy at the equilibrium lattice spacing $a_0$:
\begin{equation}
C_{ijkl} \equiv \left. \frac{\partial^2 \mathcal{F}_{\mathrm{ref}}}{\partial u_{ij}\partial u_{kl}} \right|_{a_0}.
\end{equation}
Physically, $\mathcal{F}_{\mathrm{ref}}$ may include contributions from hard-core or steric repulsion, polymer or DNA-mediated interactions, short-range solvation forces, and any external constraints that fix the system at a finite density. 
{Here $a$ denotes the radius of the equivalent spherical Wigner-Seitz cell, which is determined by the particle density; the true lattice constant of the crystal is proportional to $a$ but depends on the specific lattice type.}
{ At a fixed volume fraction (or fixed particle density $n_p = 3/(4\pi a^3)$), the Wigner-Seitz radius is fixed geometrically, while the internal ionic distribution and mechanical stresses equilibrate subject to that constraint.}
{ Purely repulsive charged colloids can crystallize at sufficiently high density or coupling strength; therefore an additional attractive mechanism is not required to stabilize the crystalline reference state.}
The stability analysis presented in this work is performed under this condition of fixed density, which is the experimentally relevant situation for colloidal crystals in a sealed sample cell.

The reference state at $a_0$ is chosen such that the total first-order stress vanishes. 
Explicitly, the total free energy expanded to quadratic order reads
\begin{equation}
\mathcal{F}_{\mathrm{tot}} = \mathcal{F}_0 + (\sigma^{\mathrm{ref}}_{ij} + \sigma^{\mathrm{es}}_{ij} + \sigma^{\mathrm{ext}}_{ij}) u_{ij} + \frac{1}{2} C_{ijkl} u_{ij} u_{kl} - \frac{\lambda_g}{2} \theta^2 + \cdots,
\end{equation}
where the linear stress terms are eliminated by the mechanical equilibrium condition $\sigma^{\mathrm{ref}}_{ij} + \sigma^{\mathrm{es}}_{ij} + \sigma^{\mathrm{ext}}_{ij}=0$. 
Here $\sigma^{\mathrm{ref}}_{ij}$ is the stress derived from the reference free energy
$\mathcal{F}_{\mathrm{ref}}$, $\sigma^{\mathrm{es}}_{ij}$ is the stress derived from
the electrostatic free energy $\mathcal{F}_{\mathrm{es}}$, and $\sigma^{\mathrm{ext}}_{ij}$ denotes the external stress arising from the boundary conditions or from the constraint that fixes the particle density. This ensures that the quadratic terms, in particular $-\frac{\lambda_g}{2} \theta^2$, represent the leading electrostatic correction to the elastic moduli without double counting the pre-stress. 
We neglect finite initial-stress corrections and assume that the reference configuration is effectively stress-free; the following stability criteria therefore apply to a mechanically equilibrated crystal with negligible Cauchy stress.

The electrostatic-elastic coupling $\lambda_g$ is obtained by expanding $\mathcal{F}_{\mathrm{es}}$ to second order in the local dilation $\theta = \nabla\cdot\mathbf{u}$. 
Because $\mathcal{F}_{\mathrm{es}}$ is not included in $\mathcal{F}_{\mathrm{ref}}$, adding the $-\frac{\lambda_g}{2}\theta^2$ term does not constitute double counting. 
The sign of $\lambda_g$ determines whether the electrostatic correction softens ($\lambda_g>0$) or stiffens ($\lambda_g<0$) the longitudinal modulus relative to the reference state. 
The spherical cell model yields an isotropic volumetric coupling, so the directional dependence of the instability arises entirely from the anisotropic background elastic constants $C_{ijkl}$, which for cubic crystals are characterized by the three independent moduli $C_{11}$, $C_{12}$, and $C_{44}$.
{The analysis that follows distinguishes carefully between the loss of strong ellipticity (acoustic phonon softening) detected by the Christoffel matrix and the thermodynamic Born stability of the crystal against homogeneous strain.}

{We stress that the elastic constants $C_{11},C_{12},C_{44}$ entering the stability criteria are the reference moduli defined from $\mathcal{F}_{\mathrm{ref}}$. 
Experimentally measured elastic moduli of charged colloidal crystals generally contain contributions from the equilibrium ionic atmosphere and may therefore differ from the bare reference values. 
A direct substitution of fully renormalized experimental moduli into our formulas would in principle require an independent determination of the reference contribution; in the absence of such a separation, the numerical examples presented in Sec.~V should be regarded as illustrations of the critical thresholds rather than as quantitative predictions for specific experimental systems.}
}

The remainder of the paper is organized as follows. 
Section~II introduces the effective static elastic tensor, the associated Christoffel matrix, and the microscopic derivation of $\lambda_g$. 
{A schematic of the spherical Wigner-Seitz cell is provided to clarify the dilation mechanism.} 
Section~III presents the directional stability criterion using the Sherman-Morrison formula. 
Section~IV evaluates the critical couplings for the $[100]$, $[110]$, and $[111]$ directions. 
Section~V discusses numerical examples, the phase diagram, the connection to unstable strain modes, and experimental implications. 
Section~VI comments on limitations and outlook, {with particular emphasis on the validity of the linearized DH approximation and its comparison with the full nonlinear PB theory.} 
A short conclusion is given in Section~VII. 
Appendix~A provides the general expression for the inverse Christoffel matrix along an arbitrary direction.

\section{Effective static elastic tensor and microscopic origin of \texorpdfstring{$\lambda_g$}{lambda\textunderscore g}}

\subsection{Elastic free energy with electrostatic-elastic coupling}

We consider a colloidal crystal that, in the absence of electrostatic-elastic coupling, is described by the reference linear elastic free energy density
\begin{equation}
\mathcal{F}_{\mathrm{ref}} = \frac{1}{2} C_{ijkl} u_{ij} u_{kl},
{\label{eq:F_el_1}}
\end{equation}
where $u_{ij} = \frac{1}{2}(\partial_i u_j + \partial_j u_i)$ is the linear strain tensor and $C_{ijkl}$ is the fourth-rank elastic modulus tensor \cite{Chaikin1995}. 
For a cubic crystal the independent non-zero components are $C_{11} = C_{xxxx}$, $C_{12} = C_{xxyy}$, and $C_{44} = C_{xyxy}$ in Voigt notation \cite{Voigt1910}.

When the crystal is immersed in an electrolyte, the electrostatic free energy $\mathcal{F}_{\mathrm{es}}$ depends on the strain through the change in the ionic environment. 
To leading order this dependence can be incorporated by adding a term proportional to $\theta^2$ in the effective free energy, reflecting the fact that a volume change alters the local screening environment and therefore the electrostatic self-energy of the lattice. 
The total static elastic free energy then takes the form
\begin{equation}
\mathcal{F}_{\mathrm{tot}} = \mathcal{F}_{\mathrm{ref}} + \Delta\mathcal{F}_{\mathrm{es}} = \int \mathrm{d}^3x \left[ \frac{1}{2} C_{ijkl} u_{ij} u_{kl} - \frac{\lambda_g}{2} (\nabla\cdot\mathbf{u})^2 \right],
\label{eq:F_eff}
\end{equation}
where $\lambda_g > 0$ is a phenomenological coupling constant with dimensions of an elastic modulus. 
The negative sign in front of $\lambda_g$ indicates that electrostatic feedback \emph{softens} the longitudinal response. 
Eq.~(\ref{eq:F_eff}) defines an effective static elastic tensor
\begin{equation}
\widetilde{C}_{ijkl} = C_{ijkl} - \lambda_g \delta_{ij} \delta_{kl}.
\end{equation}
The corresponding Christoffel matrix in Fourier space is obtained by contracting with wave vectors $k_i$:
\begin{equation}
\widetilde{M}_{ik}(\mathbf{k}) = \widetilde{C}_{ijkl} k_j k_l = M_{ik}(\mathbf{k}) - \lambda_g k_i k_k,
\label{eq:M_tilde}
\end{equation}
where $M_{ik}(\mathbf{k}) = C_{ijkl} k_j k_l$ is the bare Christoffel matrix. 
{In the harmonic lattice-dynamics approximation, the Christoffel matrix determines the long-wavelength acoustic phonon frequencies via $\rho \omega^2 e_i = M_{ik}(\mathbf{k}) e_k$, where $\rho$ is the mass density and $\mathbf{e}$ is the polarization vector; therefore, the positivity of the eigenvalues of $M(\mathbf{k})$ ensures real phonon frequencies and the loss of strong ellipticity.} 
{The analysis is restricted to the static limit; the effects of ionic diffusion on the frequency-dependent elastic moduli, though physically important for the dynamics, lie beyond the scope of the present work.}

\begin{figure}[ht]
\centering
\begin{tikzpicture}[scale=1.2]
	\draw[thick] (0,0) circle (2.0);
	\filldraw[fill=gray!30,draw=black,thick] (0,0) circle (0.8);
	\draw (0,0) node {$\bullet$};
	\draw[|<->|, thick] (0,0) -- (0.8,0) node[midway, below] {$R$};
	\draw[|<->|, thick] (0,0) -- (2.0,0) node[midway, above] {$a$};
	\node at (0.04,1.4) {Wigner-Seitz cell};
	\node at (0,-2.4) {$\partial_r\psi(a)=0$};
	\draw[->, thick, dashed] (2.2,0) -- (2.8,0) node[right] {$a \to a(1+\theta/3)$};
\end{tikzpicture}
\caption{Schematic of the spherical Wigner-Seitz cell. A colloidal particle of radius $R$ (grey) is surrounded by an electrolyte-filled cell of radius $a$. The cell is charge neutral, so the electric field vanishes at the outer boundary. A local dilation $\theta = \nabla\cdot\mathbf{u}$ changes the cell radius to $a(1+\theta/3)$. This volume change alters the ionic distribution and gives rise to the electrostatic-elastic coupling.}
\label{fig:cell}
\end{figure}

\subsection{Microscopic derivation of \texorpdfstring{$\lambda_g$}{lambda\textunderscore g} from the Poisson-Boltzmann cell model}
\label{sec:micro_derivation}

The coupling constant $\lambda_g$ has so far been introduced phenomenologically. 
To connect it with experimentally accessible parameters, we derive its expression from the nonlinear PB theory in a spherical Wigner-Seitz cell approximation \cite{Alexander1984, Belloni1998, Safran1994}. 
{A schematic of the cell is shown in Fig.~\ref{fig:cell}.} 
{Consider a colloidal crystal with spherical particles of radius $R$, each centred in a spherical Wigner-Seitz cell of radius $a$.}
{
Each particle is located at the centre of a spherical cell of radius $a$, within which the electrostatic potential $\psi(r)$ satisfies the PB equation. 
The cell is overall neutral, implying $\partial_r \psi(a)=0$. 
The electrolyte is a monovalent (1:1) salt with bulk ion concentration $n_0$. 
The local ionic number densities are $n_{\pm}(\mathbf{r}) = n_0\, e^{\mp e_0\psi(\mathbf{r})/k_B T}$ and the charge density is $\rho(\mathbf{r}) = e_0[n_+(\mathbf{r}) - n_-(\mathbf{r})]$, representing the mobile-ion charge density only. 
The fixed surface charge on the colloidal particle is treated through the boundary condition.
It is convenient to introduce the dimensionless potential $y \equiv e_0 \psi / k_B T$, which obeys
}
\begin{equation}
\nabla^2 y = \kappa_0^2 \sinh y,
\label{eq:PB_dimensionless}
\end{equation}
where $\kappa_0^2 = 8\pi \ell_B n_0$ is the bare Debye screening constant, $\ell_B = e_0^2/(4\pi\epsilon k_B T)$ is the Bjerrum length, $\epsilon$ is the dielectric permittivity of the solvent, and $k_B T$ is the thermal energy. 
The boundary conditions are $y'(a)=0$ (cell neutrality) and $-\epsilon y'(R) = \sigma e_0 / k_B T$ (constant surface charge). 
{
The PB equation is a mean-field approximation that neglects microion correlations. 
The subsequent linearization to the DH limit requires that the dimensionless electrostatic potential be small throughout the cell, $|y(r)| \ll 1$. 
This condition imposes constraints on the surface charge density and the ionic strength that must be verified for the parameter range of interest. 
Additionally, the mean-field nature of PB theory requires weak ion-ion correlations, which is satisfied when $\ell_B \ll \kappa_0^{-1}$ and $\ell_B \ll R$.
}

{
The electrostatic free energy of the cell (per particle) consists of two physically distinct contributions \cite{Levin2002}: 
the electrostatic self-energy of the complete charge distribution (mobile ions plus fixed surface charge)
\begin{equation}
\frac{1}{2} \int_{\mathrm{cell}} \rho_{\mathrm{tot}}(\mathbf{r}) \psi(\mathbf{r}) \, \mathrm{d}^3r,
\end{equation}
and the entropic free energy of mixing of the mobile ions
\begin{equation}
\begin{aligned}
&k_B T \int_{\mathrm{cell}} \Big[ n_+(\mathbf{r}) \ln\frac{n_+(\mathbf{r})}{n_0} + n_-(\mathbf{r}) \ln\frac{n_-(\mathbf{r})}{n_0} \\
&- (n_+(\mathbf{r}) + n_-(\mathbf{r}) - 2n_0) \Big] \mathrm{d}^3r.
\end{aligned}
\end{equation}
The total free energy per cell is thus
\begin{equation}
\begin{aligned}
F_{\mathrm{es}}(a) =& \frac{1}{2} \int_{\mathrm{cell}} \rho_{\mathrm{tot}}(\mathbf{r}) \psi(\mathbf{r}) \, \mathrm{d}^3r + k_B T \int_{\mathrm{cell}} \Big[ n_+(\mathbf{r}) \ln\frac{n_+(\mathbf{r})}{n_0} \\
&+ n_-(\mathbf{r}) \ln\frac{n_-(\mathbf{r})}{n_0} - (n_+(\mathbf{r}) + n_-(\mathbf{r}) - 2n_0) \Big] \mathrm{d}^3r.
\label{eq:Fes}
\end{aligned}
\end{equation}
Although the variable $a$ does not appear explicitly in the integrand of Eq.~\eqref{eq:Fes}, 
the free energy depends on $a$ implicitly because the electrostatic potential $\psi(\mathbf{r})$ 
and the ionic densities $n_{\pm}(\mathbf{r})$ are obtained by solving the Poisson--Boltzmann 
equation inside a spherical cell of radius $a$ with the boundary condition 
$\partial_r \psi(a)=0$ [Eq.~\eqref{eq:PB_dimensionless}]. 
A change in the cell radius therefore modifies the solution of the Poisson--Boltzmann 
equation and, consequently, the value of $F_{\mathrm{es}}$.
}
{In the DH limit, the cell is assumed to be in contact with a salt reservoir that fixes the chemical potential of the ions, so that the Debye parameter $\kappa_0$ is held constant during the deformation.}
For small dilations, { the Wigner-Seitz cell radius} changes as $a \to a(1+\theta/3)$, { reflecting the isotropic volume change of a cubic cell.} 
The corresponding change in cell radius is $\Delta a = a\theta/3$ {(see Fig.~\ref{fig:cell})}. 
{ The crystal is assumed to be held at a fixed particle density (e.g., in a sealed sample cell), so that the Wigner-Seitz radius $a_0$ is a predetermined experimental parameter. The internal degrees of freedom (ionic distribution, mechanical stresses) then equilibrate at this fixed cell volume.} 
Expanding $F_{\mathrm{es}}$ around $a_0$ to second order in $\Delta a$ gives
\begin{equation}
\Delta F_{\mathrm{es}} \approx \frac{1}{2} \frac{\partial^2 F_{\mathrm{es}}}{\partial a^2} (\Delta a)^2 
= \frac{a^2}{18} \frac{\partial^2 F_{\mathrm{es}}}{\partial a^2} \theta^2.
\end{equation}
{Because $F_{\mathrm{es}}$ is defined as the part of the total free energy that is not contained in $\mathcal{F}_{\mathrm{ref}}$, and because the reference state at $a_0$ is chosen such that the total first-order stress vanishes (see the explicit cancellation of linear terms in Section~I), the quadratic expansion of $\Delta\mathcal{F}_{\mathrm{es}}$ alone correctly captures the leading electrostatic correction to the elastic moduli.}
Dividing by the cell volume $V_c = \frac{4\pi}{3} a^3$ yields the change in free energy per unit volume,
\begin{equation}
\Delta \mathcal{F}_{\mathrm{es}} \approx \frac{a^2}{18 V_c} \frac{\partial^2 F_{\mathrm{es}}}{\partial a^2} \theta^2.
\label{eq:deltaF}
\end{equation}
Comparing Eq.~(\ref{eq:deltaF}) with the continuum expression $-\frac{\lambda_g}{2} \theta^2$ from Eq.~\eqref{eq:F_eff}, we identify
\begin{equation}
\lambda_g = -\frac{a^2}{9 V_c} \frac{\partial^2 F_{\mathrm{es}}}{\partial a^2}.
\label{eq:lambda_def}
\end{equation}
{The promotion of the cell free-energy curvature to a local continuum coupling $-\frac{\lambda_g}{2}[\nabla\cdot\mathbf{u}(\mathbf{x})]^2$ implicitly assumes that the deformation wavelength is large compared to both the cell size $a$ and the screening length $\kappa_0^{-1}$, so that the local volume change can be treated as a homogeneous dilation of the cell.}
The second derivative $\partial^2 F_{\mathrm{es}}/\partial a^2$ can be evaluated numerically for given $R$, $\sigma$, and the dimensionless screening parameter $\kappa_0 a$. 
In the Debye-H\"uckel (linear) limit, the PB equation can be solved analytically, yielding \cite{Russell1989}
\begin{equation}
\lambda_g^{\mathrm{DH}} = \frac{Z^2 e_0^2}{48\pi \epsilon a^4} \frac{\kappa_0 a \cosh(\kappa_0 a) - \sinh(\kappa_0 a)}{(\kappa_0 a)^3 \sinh^2(\kappa_0 a)}.
\label{eq:lambda_DH}
\end{equation}
{ A direct inspection shows that $\lambda_g^{\mathrm{DH}} > 0$ for all physically relevant $\kappa_0 a > 0$ (the numerator is always positive).} 
Thus, even in the linearized DH limit, electrostatic-elastic coupling predicts a softening of the longitudinal acoustic modes. 
{
For the parameter range relevant to the experiments discussed in Sec.~V 
($Z \sim 10^2$--$10^3$, $a \sim 1\,\mu\text{m}$, $\kappa_0 a \sim 1$--$3$), 
$\lambda_g$ falls in the range $1$--$10\,k_B T/\mu\text{m}^3$, consistent with 
the numerical values adopted.
}
{
It is crucial to emphasize that Eq.~\eqref{eq:lambda_DH} results from a \emph{linearization} 
of the nonlinear PB functional. Tamashiro and Schiessel~\cite{Tamashiro2003} have rigorously 
analysed this linearization procedure in the spherical cell model. They demonstrated that the 
linearized theory is asymptotically exact only in the weak-coupling (high-temperature) limit, 
and becomes unreliable at strong electrostatic coupling, where it can produce spurious phase 
separation and negative compressibilities. 

For a representative set of experimentally relevant parameters ($Z = 500$, $a \approx 1\,\mu$m, 
$\kappa_0 a \approx 2$), the electrostatic coupling parameter is 
$\Xi \equiv Z\ell_B/[a(\kappa_0 a)^2] \approx 0.09 \ll 1$, and the effective charge parameter 
$\theta \equiv 3Z\ell_B/a \approx 1.05$ is far below the critical value $\theta_{\text{crit}} \approx 45$ 
found by Tamashiro and Schiessel for the onset of spurious instabilities. 
{For the same parameters, the dimensionless surface potential estimated from the DH solution is $|y(R)| \approx 0.5$, confirming that the linearization condition $|y(r)|\ll 1$ is well satisfied.}
These estimates place our system firmly in the weak-coupling regime, confirming that the 
linearized DH approximation is well justified. 
Our analytical expression, within its range of validity, therefore offers a transparent link 
between the phenomenological coupling and the experimental control parameters, capturing the 
softening trend without the artifacts that plague the linearized theory in the strong-coupling 
regime.
}

We also note that the derivation assumes isotropic cell deformation; for anisotropic crystals, the coupling may acquire direction dependence, which is left for future investigation. 
The spherical cell approximation neglects the faceted shape of the true Wigner-Seitz cell. 
For dilute colloidal crystals (volume fraction $\lesssim 10^{-3}$) this approximation introduces an error of less than $10\%$ in the electrostatic free energy~\cite{Alexander1984, Robbins1988}. 
{ Such extremely dilute crystals are experimentally realizable because the long range of electrostatic repulsion can maintain crystalline order even at very low volume fractions; for instance, Reus \emph{et al.}~\cite{reus1997equation} observed stable bcc crystals at volume fractions as low as $0.04\%$ (see also Ref.~\cite{Sood1991}).} 
At higher densities, a full numerical solution of the PB equation in the actual periodic geometry would be required.

\section{Directional stability criterion}

Because $\widetilde{M}(\mathbf{k})$ is a real symmetric $3\times3$ matrix, it is positive definite if and only if all its eigenvalues are positive. 
The softening term $-\lambda_g \mathbf{k}\otimes\mathbf{k}$ is a rank-one perturbation of $M(\mathbf{k})$. 
A compact stability condition can therefore be obtained using the matrix determinant lemma.

We write
\begin{equation}
\widetilde{M}(\mathbf{k}) = M(\mathbf{k}) \left[ I - \lambda_g M^{-1}(\mathbf{k}) (\mathbf{k}\otimes\mathbf{k}) \right].
\end{equation}
The determinant is
\begin{equation}
\det \widetilde{M}(\mathbf{k}) = \det M(\mathbf{k}) \cdot \det\left[ I - \lambda_g M^{-1}(\mathbf{k}) \mathbf{k}\mathbf{k}^\mathrm{T} \right].
\end{equation}
For a rank-one matrix $\mathbf{u}\mathbf{v}^\mathrm{T}$, we have $\det(I - \mathbf{u}\mathbf{v}^\mathrm{T}) = 1 - \mathbf{v}^\mathrm{T} \mathbf{u}$. 
Applying this identity with $\mathbf{u} = \lambda_g M^{-1}\mathbf{k}$ and $\mathbf{v} = \mathbf{k}$ yields
\begin{equation}
\det \widetilde{M}(\mathbf{k}) = \det M(\mathbf{k}) \left[ 1 - \lambda_g \, \mathbf{k}^\mathrm{T} M^{-1}(\mathbf{k}) \mathbf{k} \right].
\label{eq:det}
\end{equation}
Since the bare elastic tensor is positive definite, $\det M(\mathbf{k}) > 0$ for all $\mathbf{k}\neq 0$. 
Therefore the sign of $\det \widetilde{M}(\mathbf{k})$ is determined by the factor in brackets. 
{A loss of strong ellipticity (vanishing acoustic phonon frequency)} occurs when
\begin{equation}
1 - \lambda_g \, \mathbf{k}^\mathrm{T} M^{-1}(\mathbf{k}) \mathbf{k} = 0.
\label{eq:instab_cond}
\end{equation}
For a given direction $\hat{\mathbf{k}} = \mathbf{k}/k$, the bare Christoffel matrix scales as $M(\mathbf{k}) = k^2 \Gamma(\hat{\mathbf{k}})$, where $\Gamma(\hat{\mathbf{k}})$ depends only on the direction. 
Consequently,
\begin{equation}
\mathbf{k}^\mathrm{T} M^{-1}(\mathbf{k}) \mathbf{k} = \hat{\mathbf{k}}^\mathrm{T} \Gamma^{-1}(\hat{\mathbf{k}}) \hat{\mathbf{k}} \equiv K(\hat{\mathbf{k}}),
\label{eq:K_def}
\end{equation}
which is independent of the magnitude $k$. 
{The acoustic stability criterion for a given propagation direction} then becomes
\begin{equation}
\lambda_g < \frac{1}{K(\hat{\mathbf{k}})} \qquad \text{for all directions } \hat{\mathbf{k}}.
\label{eq:stab_crit}
\end{equation}
{The critical coupling for the loss of strong ellipticity along a specific direction} is
\begin{equation}
\lambda_g^{c}(\hat{\mathbf{k}}) = \frac{1}{K(\hat{\mathbf{k}})}.
\end{equation}
{The direction whose acoustic branch first softens as $\lambda_g$ increases is the one that minimizes $\lambda_g^{c}(\hat{\mathbf{k}})$, i.e., the direction that maximizes $K(\hat{\mathbf{k}})$.}

It is instructive to note that $K(\hat{\mathbf{k}})$ is exactly the static longitudinal compliance of the bare elastic medium along direction $\hat{\mathbf{k}}$. 
Indeed, if we apply a uniaxial stress that generates a longitudinal strain wave with wave vector $\mathbf{k}$, the elastic energy density is $\frac{1}{2} (\hat{k}_i \Gamma_{ik} \hat{k}_k) |\tilde{\theta}|^2$, and $1/K(\hat{\mathbf{k}})$ is the corresponding longitudinal modulus. 
The electrostatic-elastic coupling subtracts $\lambda_g$ from this longitudinal modulus, and {an acoustic} instability occurs when the renormalized modulus vanishes.

\subsection{{Born stability against homogeneous strain}}
\label{sec:Born}
{
The Christoffel criterion derived above is a condition for the loss of strong ellipticity, which is a necessary but not sufficient condition for the total elastic free energy to be bounded from below against all homogeneous strains. 
For the effective elastic tensor $\widetilde{C}_{ijkl} = C_{ijkl} - \lambda_g \delta_{ij} \delta_{kl}$, the thermodynamic Born stability conditions for a cubic crystal are:
\begin{align}
\widetilde{C}_{44} &> 0, \notag\\
\widetilde{C}_{11} - \widetilde{C}_{12} &> 0, \\
\widetilde{C}_{11} + 2\widetilde{C}_{12} &> 0. \notag
\end{align}
Substituting the effective moduli yields:
\begin{align}
C_{44} &> 0, \notag\\
C_{11} - C_{12} &> 0, \\
C_{11} + 2C_{12} - 3\lambda_g &> 0. \notag
\end{align}
The first two conditions are independent of $\lambda_g$ and are satisfied for any stable bare crystal. 
The third condition gives the critical coupling for a homogeneous volumetric (Born) instability:
\begin{equation}
\lambda_g^{\mathrm{Born}} = \frac{C_{11} + 2C_{12}}{3} = K, \label{eq:lambda_Born}
\end{equation}
where $K$ is the bare bulk modulus. 
This threshold is isotropic and independent of the propagation direction. 
{For a Born-stable reference cubic crystal, the homogeneous volume mode becomes unstable before any of the three principal acoustic thresholds, provided homogeneous dilation is an admissible deformation.}

A direct comparison shows that the acoustic thresholds derived from the Christoffel matrix are always greater than or equal to the Born threshold:
\begin{align}
\lambda_g^{c}([100]) - \lambda_g^{\mathrm{Born}} &= \frac{2}{3}(C_{11} - C_{12}) > 0, \\
\lambda_g^{c}([111]) - \lambda_g^{\mathrm{Born}} &= \frac{4}{3}C_{44} > 0.
\end{align}
Therefore, in a system where uniform volume changes are allowed, the thermodynamic Born instability occurs at a lower critical coupling than any directional acoustic instability derived from the Christoffel analysis above. 
The acoustic thresholds and the associated direction-dependent phase diagram (Figs.~\ref{fig:anisotropy}--\ref{fig:phasediagram}) are physically relevant when the total volume of the crystal is constrained, for instance in a sealed sample cell with fixed particle density, which is the scenario assumed in the present work. 
{Under such a strict fixed-volume constraint, the homogeneous $k=0$ dilation mode is excluded from the admissible perturbations, while finite-wave-vector longitudinal modes with zero spatially averaged dilation remain allowed.}
In this constrained case, where homogeneous volume changes are suppressed, the Christoffel analysis correctly identifies the first directional acoustic mode to soften.
}

\section{Explicit evaluation for cubic symmetry}

For a cubic crystal, the Christoffel matrix $\Gamma(\hat{\mathbf{k}})$ has components
\begin{equation}
\begin{aligned}
\Gamma_{11} &= C_{11} \hat{k}_1^2 + C_{44} (\hat{k}_2^2 + \hat{k}_3^2), \\
\Gamma_{22} &= C_{11} \hat{k}_2^2 + C_{44} (\hat{k}_1^2 + \hat{k}_3^2), \\
\Gamma_{33} &= C_{11} \hat{k}_3^2 + C_{44} (\hat{k}_1^2 + \hat{k}_2^2), \\
\Gamma_{12} = \Gamma_{21} &= (C_{12} + C_{44}) \hat{k}_1 \hat{k}_2, \\
\Gamma_{13} = \Gamma_{31} &= (C_{12} + C_{44}) \hat{k}_1 \hat{k}_3, \\
\Gamma_{23} = \Gamma_{32} &= (C_{12} + C_{44}) \hat{k}_2 \hat{k}_3.
\end{aligned}
\label{eq:Gamma_comp}
\end{equation}
While the full angular dependence of $K(\hat{\mathbf{k}})$ can be studied numerically (the general algebraic form is provided in Appendix~A, the analysis simplifies greatly along the three high-symmetry directions where $\hat{\mathbf{k}}$ is an eigenvector of $\Gamma(\hat{\mathbf{k}})$.

\subsection{Direction \texorpdfstring{$[100]$}{[100]}}
For $\hat{\mathbf{k}} = (1,0,0)$, the Christoffel matrix is diagonal:
\[
\Gamma([100]) = \begin{pmatrix}
C_{11} & 0 & 0 \\
0 & C_{44} & 0 \\
0 & 0 & C_{44}
\end{pmatrix}.
\]
Hence $K([100]) = 1/C_{11}$ and
\begin{equation}
\lambda_g^{c}([100]) = C_{11}.
\label{eq:lam100}
\end{equation}

\subsection{Direction \texorpdfstring{$[110]$}{[110]}}
For $\hat{\mathbf{k}} = \frac{1}{\sqrt{2}}(1,1,0)$,
\[
\Gamma([110]) = \begin{pmatrix}
\frac{C_{11}+C_{44}}{2} & \frac{C_{12}+C_{44}}{2} & 0 \\[4pt]
\frac{C_{12}+C_{44}}{2} & \frac{C_{11}+C_{44}}{2} & 0 \\[4pt]
0 & 0 & C_{44}
\end{pmatrix}.
\]
The eigenvector $\frac{1}{\sqrt{2}}(1,1,0)$ has eigenvalue
\[
\lambda_L = \frac{C_{11} + C_{12} + 2C_{44}}{2}.
\]
Therefore $K([110]) = 2/(C_{11}+C_{12}+2C_{44})$ and
\begin{equation}
\lambda_g^{c}([110]) = \frac{C_{11} + C_{12} + 2C_{44}}{2}.
\label{eq:lam110}
\end{equation}

\subsection{Direction \texorpdfstring{$[111]$}{[111]}}
For $\hat{\mathbf{k}} = \frac{1}{\sqrt{3}}(1,1,1)$,
\[
\Gamma([111]) = \begin{pmatrix}
a & b & b \\
b & a & b \\
b & b & a
\end{pmatrix}, \
a = \frac{C_{11}+2C_{44}}{3}, \ b = \frac{C_{12}+C_{44}}{3}.
\]
The eigenvector $\frac{1}{\sqrt{3}}(1,1,1)$ has eigenvalue
\[
\lambda_L = a + 2b = \frac{C_{11} + 2C_{12} + 4C_{44}}{3}.
\]
Hence $K([111]) = 3/(C_{11}+2C_{12}+4C_{44})$ and
\begin{equation}
\lambda_g^{c}([111]) = \frac{C_{11} + 2C_{12} + 4C_{44}}{3}.
\label{eq:lam111}
\end{equation}

\begin{figure*}[htbp]
\centering
\includegraphics[width=0.95\textwidth]{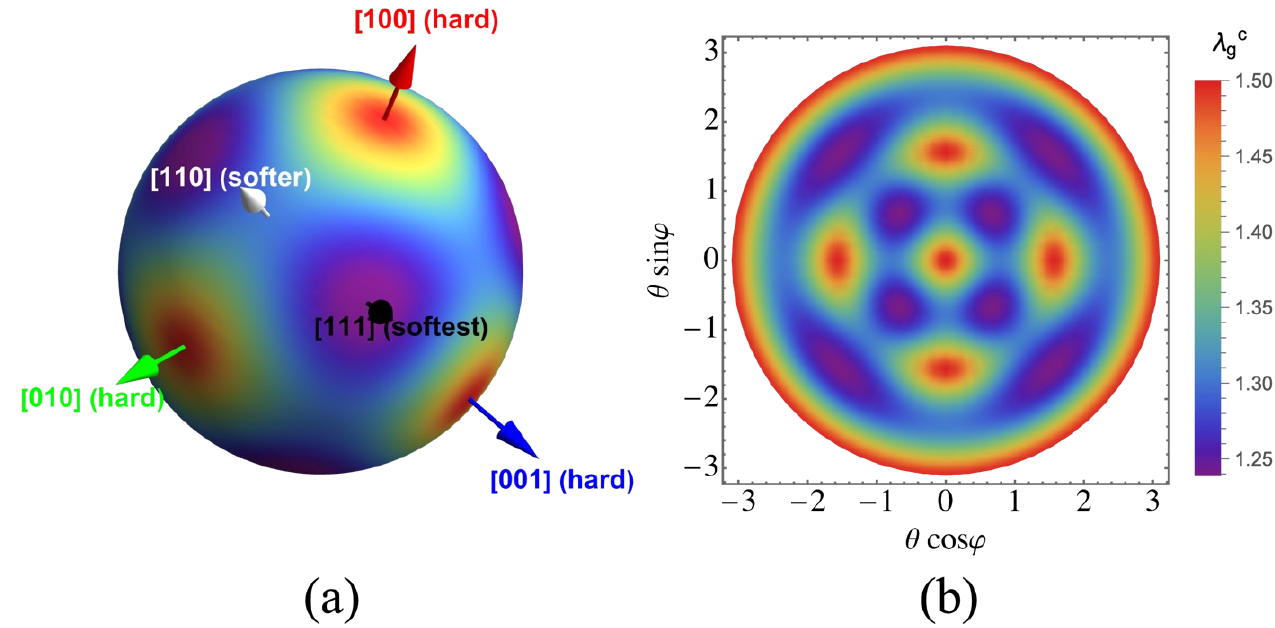}
\caption{Directional anisotropy of the critical electrostatic-elastic coupling $\lambda_g^c$ for a cubic colloidal crystal with $C_{11}=1.5$, $C_{12}=0.5$, $C_{44}=0.3$. (a) Spherical color map showing the full solid-angle dependence. The color scale ranges from blue (minimum $\lambda_g^c \approx 1.233$) to red (maximum $\lambda_g^c = 1.50$). Solid red, green, and blue arrows indicate the $[100]$, $[010]$, and $[001]$ hard directions; white and black arrows mark the $[110]$ and $[111]$ directions, respectively. (b) Projection onto the $(\theta,\phi)$ plane, which avoids the area distortion of stereographic projection. In both panels the minimum of $\lambda_g^c$ occurs along the $[111]$ family (body-diagonal directions), while the $[110]$ directions exhibit a slightly higher value ($\approx 1.30$).}
\label{fig:anisotropy}
\end{figure*}

\begin{table}[h]
\centering
\caption{Critical electrostatic-elastic coupling $\lambda_g^{c}$ along high-symmetry directions in a cubic crystal.}
\label{tab:critical}
\begin{tabular}{@{}ll@{}}
\toprule
Direction & $\lambda_g^{c}$ \\
\midrule
$[100]$ & $C_{11}$ \\[4pt]
$[110]$ & $\dfrac{C_{11} + C_{12} + 2C_{44}}{2}$ \\[8pt]
$[111]$ & $\dfrac{C_{11} + 2C_{12} + 4C_{44}}{3}$ \\
\bottomrule
\end{tabular}
\end{table}

The three critical couplings are summarized in Table~\ref{tab:critical}. 
In the isotropic limit $C_{11} = \lambda+2\mu$, $C_{12} = \lambda$, $C_{44} = \mu$, all three expressions reduce to $\lambda+2\mu$, recovering the isotropic result of Refs.~\cite{Wu2025,Wu2024EPL} and the classic Lam\'e theory \cite{Landau1986}. 
This confirms that our anisotropic formulas correctly degenerate to the known isotropic case.

For later reference, we collect the three expressions in a single equation:
\begin{align}
\lambda_g^{c}([100]) &= C_{11}, \notag\\
\lambda_g^{c}([110]) &= \frac{C_{11} + C_{12} + 2C_{44}}{2}, \\
\lambda_g^{c}([111]) &= \frac{C_{11} + 2C_{12} + 4C_{44}}{3}. \notag
\label{eq:lam3}
\end{align}

\section{Numerical illustrations and discussion}

The critical coupling depends on the three cubic moduli in a non-trivial way. 
To illustrate the directional ordering we consider two representative parameter sets taken from experiments on soft colloidal crystals.

\subsection{Example 1: bcc colloidal crystal of charged polystyrene spheres}
Dubois-Violette \emph{et al.} measured the elastic constants of a bcc colloidal crystal formed by charged polystyrene spheres in deionized water \cite{Dubois1980}. 
{We use their reported values as representative reference moduli solely to illustrate the threshold behaviour; they are not claimed to be the bare reference constants $C_{ijkl}$ of the present theory.}
Typical values (in units of $k_BT/d^3$, where $d$ is the mean particle spacing) are $C_{11} \approx 1.5$, $C_{12} \approx 0.5$, $C_{44} \approx 0.3$. 
Inserting these numbers into Eq.~(\ref{eq:lam3}) yields
\begin{align*}
\lambda_g^{c}([100]) &= 1.50, \\
\lambda_g^{c}([110]) &= 1.30, \\
\lambda_g^{c}([111]) &\approx 1.233.
\end{align*}
The smallest threshold is $\lambda_g^{c}([111])$, indicating that the body-diagonal direction softens first for this parameter set. 
This result is consistent with the general ordering relation derived below: for $\Delta = C_{11}-C_{12}-2C_{44} > 0$, the sequence is $\lambda_g^c([100]) > \lambda_g^c([110]) > \lambda_g^c([111])$, so $[111]$ is the softest.

\begin{figure}[ht]
\centering
\includegraphics[width=1.0\columnwidth]{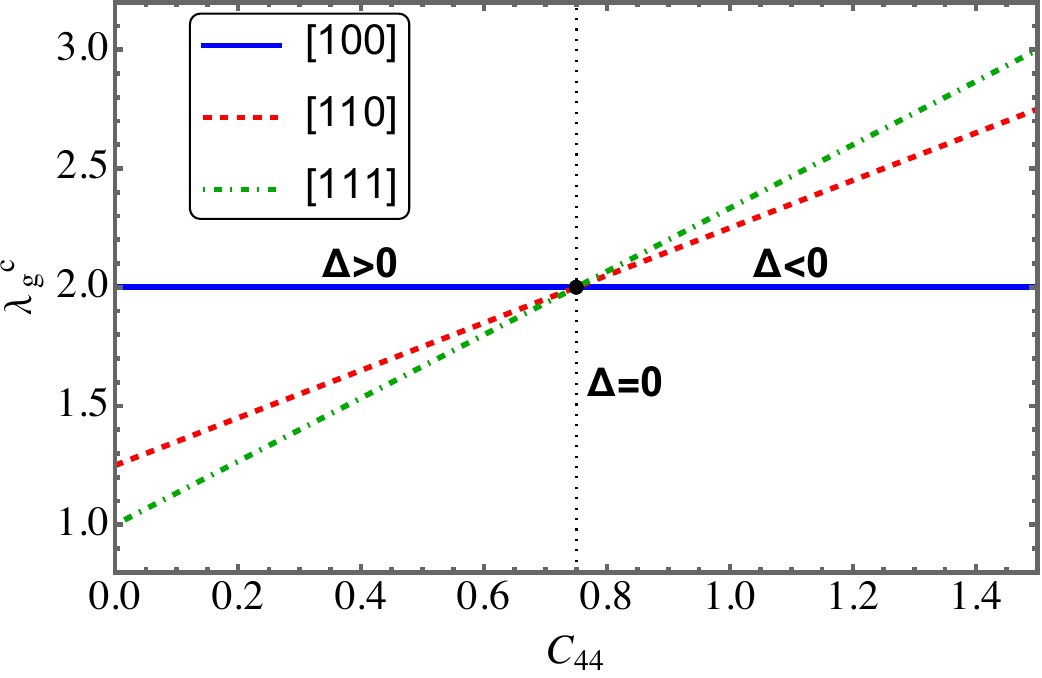}
\caption{Dependence of the critical coupling $\lambda_g^{c}$ on the shear modulus $C_{44}$ for fixed $C_{11}=2.0$, $C_{12}=0.5$. The curves intersect at $C_{44} = (C_{11}-C_{12})/2 = 0.75$, corresponding to the condition $\Delta \equiv C_{11} - C_{12} - 2C_{44} = 0$. To the left of the intersection ($\Delta > 0$), the ordering is $\lambda_g^c([100]) > \lambda_g^c([110]) > \lambda_g^c([111])$, so the body-diagonal $[111]$ direction is the softest. To the right of the intersection ($\Delta < 0$), the ordering reverses to $\lambda_g^c([111]) > \lambda_g^c([110]) > \lambda_g^c([100])$, making the cube-edge $[100]$ direction the softest. The face-diagonal $[110]$ direction is always intermediate between the other two.}
\label{fig:C44}
\end{figure}

To gain a more complete picture of the angular dependence of $\lambda_g^c$ beyond the three high-symmetry directions, we visualize its full solid-angle variation in Fig.~\ref{fig:anisotropy}. 
Panel (a) displays a three-dimensional spherical color map for the same elastic constants. 
The color scale ranges from the minimum value $\approx 1.233$ (blue) to the maximum $\approx 1.50$ (red). 
To facilitate orientation, we have added arrows indicating the principal crystallographic axes: the cube edges $[100]$, $[010]$, $[001]$ are shown as solid red, green, and blue lines, respectively, and coincide with the red regions, confirming that these directions are the hardest against electrostatic-elastic softening. 
In contrast, the dashed black arrows point along $[110]$ and $[111]$; the $[111]$ direction lies in the deep blue region, while the $[110]$ direction appears in the blue-green transition, confirming the ordering of thresholds. 
Panel (b) provides an alternative representation in polar coordinates $(\theta,\phi)$, where the radial coordinate is the polar angle $\theta$ and the azimuthal angle is $\phi$. 
This projection preserves the area element of the sphere and avoids the peripheral distortion inherent in stereographic projections. 
The centre of the plot corresponds to the $[001]$ pole, while the outer circular boundary corresponds to the opposite pole $[00\bar{1}]$. 
The four-fold symmetric lobes of low $\lambda_g^c$ around $\theta=\pi/2$ correspond to the $[111]$ family of soft directions. 
The agreement between the two panels demonstrates that the anisotropy pattern is robust and dominated by the relatively small shear modulus $C_{44}$.

\subsection{Example 2: DNA-grafted nanoparticle superlattices}
DNA-mediated colloidal crystals often exhibit a more central-force character, with $C_{12} \approx C_{44}$ \cite{Park2008, Macfarlane2011}. 
For instance, using $C_{11}=2.0$, $C_{12}=0.8$, $C_{44}=0.5$ we obtain
\begin{align*}
\lambda_g^{c}([100]) &= 2.00, \\
\lambda_g^{c}([110]) &= 1.90, \\
\lambda_g^{c}([111]) &= 1.87.
\end{align*}
Here the $[111]$ direction is marginally the softest. 
The differences between the three values are, however, quite small; the crystal is nearly isotropic in its longitudinal response.

\subsection{Influence of the anisotropy ratio}

The degree of elastic anisotropy is often quantified by the Zener ratio $A = 2C_{44}/(C_{11}-C_{12})$ \cite{Zener1948}. 
For an isotropic medium $A=1$. 
In Fig.~\ref{fig:C44} we plot the variation of the three critical couplings as a function of $C_{44}$ while keeping $C_{11}$ and $C_{12}$ fixed at $C_{11}=2.0$ and $C_{12}=0.5$. 
For small $C_{44}$ the $[110]$ and $[111]$ directions are significantly softer than $[100]$; as $C_{44}$ increases, all three thresholds approach each other and eventually cross at the point where $\Delta \equiv C_{11} - C_{12} - 2C_{44} = 0$. 
This intersection marks the boundary between two distinct regimes: 
\begin{itemize}
    \item For $\Delta > 0$ (left side of the intersection), the ordering is $\lambda_g^c([100]) > \lambda_g^c([110]) > \lambda_g^c([111])$, so the body-diagonal $[111]$ direction is the first to soften.
    \item For $\Delta < 0$ (right side of the intersection), the ordering reverses to $\lambda_g^c([111]) > \lambda_g^c([110]) > \lambda_g^c([100])$, making the cube-edge $[100]$ direction the softest.
\end{itemize}
Remarkably, the face-diagonal $[110]$ direction is never the unique minimum; it is always intermediate between the other two critical values. 
The phase diagram shown in Fig.~\ref{fig:phasediagram} provides a complete view of this behavior in the full parameter space spanned by the reduced elastic constants $C_{44}/C_{11}$ and $C_{12}/C_{11}$.

\subsection{Phase diagram in the space of elastic constants}

\begin{figure}[ht]
\centering
\includegraphics[width=\columnwidth]{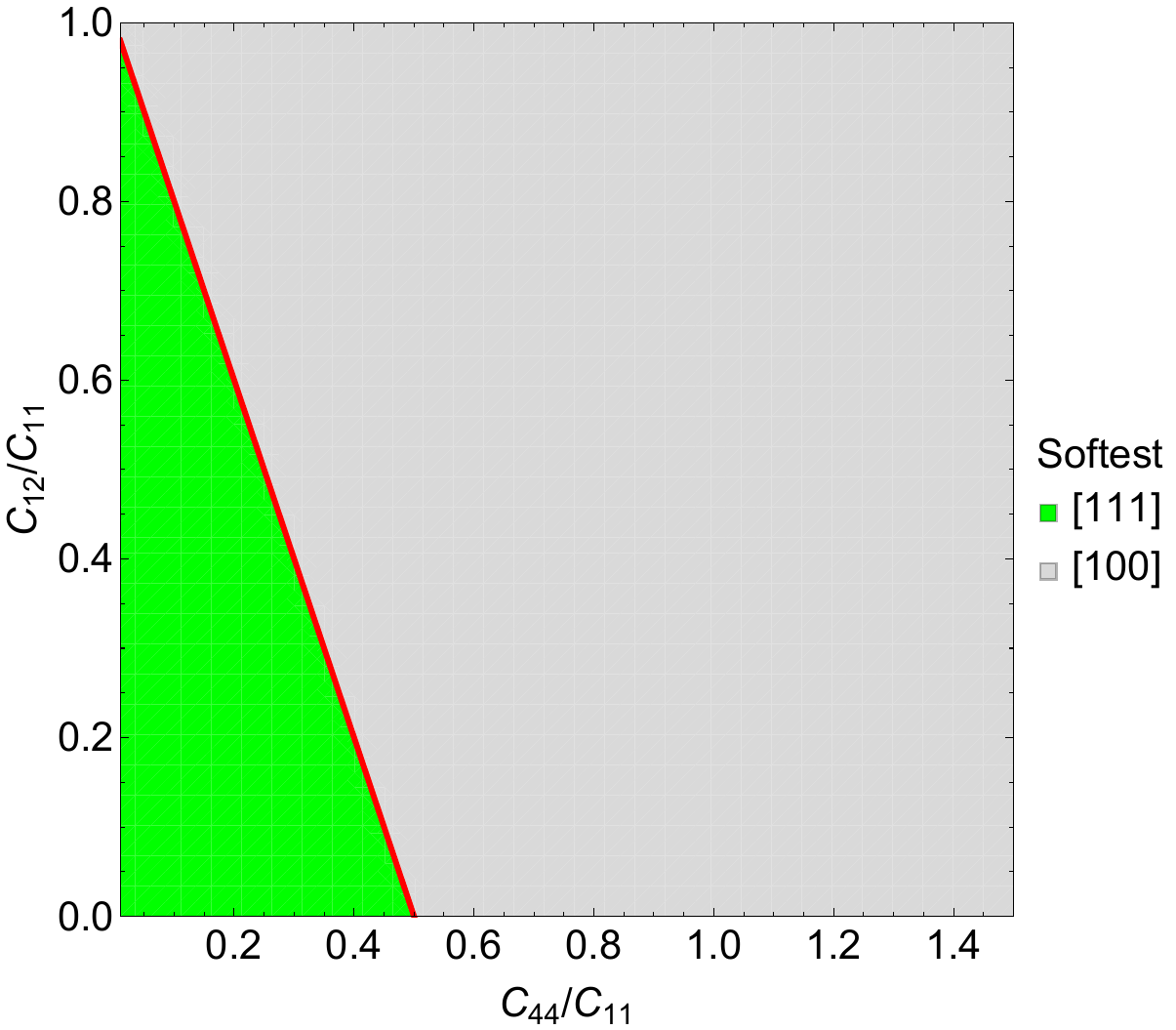}
\caption{Phase diagram of the softest crystallographic direction in a cubic colloidal crystal as a function of the reduced elastic constants $C_{44}/C_{11}$ and $C_{12}/C_{11}$. The green region ($\Delta > 0$) corresponds to the body-diagonal $[111]$ direction being the softest, while the gray region ($\Delta < 0$) corresponds to the cube-edge $[100]$ direction being the softest. The face-diagonal $[110]$ direction is never the unique softest direction under scalar electrostatic-elastic coupling; it is always intermediate except on the red boundary line $C_{11} - C_{12} - 2C_{44} = 0$, where all three directions become degenerate.}
\label{fig:phasediagram}
\end{figure}

As derived in Sec.~IV, the three critical couplings satisfy the ordering relations
\begin{align}
\lambda_g^{c}([100]) - \lambda_g^{c}([110]) &= \frac{\Delta}{2}, \nonumber \\
\lambda_g^{c}([100]) - \lambda_g^{c}([111]) &= \frac{2\Delta}{3}, \\
\lambda_g^{c}([110]) - \lambda_g^{c}([111]) &= \frac{\Delta}{6}, \nonumber
\label{eq:diff}
\end{align}
with $\Delta \equiv C_{11} - C_{12} - 2C_{44}$. 
From these relations we deduce the exact ordering of the critical couplings:
\begin{itemize}
    \item If $\Delta > 0$: $\lambda_g^c([100]) > \lambda_g^c([110]) > \lambda_g^c([111])$, so the body-diagonal $[111]$ direction is the softest.
    \item If $\Delta < 0$: $\lambda_g^c([100]) < \lambda_g^c([110]) < \lambda_g^c([111])$, so the cube-edge $[100]$ direction is the softest.
    \item If $\Delta = 0$: all three critical couplings are equal, corresponding to an effectively isotropic longitudinal response.
\end{itemize}
A remarkable consequence of this analysis is that the face-diagonal direction $[110]$ never constitutes the unique softest direction; it is always the intermediate value except at the degenerate boundary. 
This counterintuitive result stems from the fact that the scalar electrostatic-elastic coupling $\lambda_g$ couples only to the trace of the strain tensor, which weights the three cubic moduli in a specific manner.
A more general analysis using the Sherman-Morrison expression for arbitrary directions (see Appendix~A) confirms that the global minimum of $\lambda_g^c(\hat{\mathbf{k}})$ always occurs at either $[100]$ or $[111]$, with $[110]$ never being the unique minimum.

Figure~\ref{fig:phasediagram} displays the resulting phase diagram in the space of reduced elastic constants $C_{44}/C_{11}$ and $C_{12}/C_{11}$. 
The green region ($\Delta > 0$) corresponds to $[111]$ being the softest, while the gray region ($\Delta < 0$) corresponds to $[100]$ being the softest. 
The red boundary line $\Delta = 0$ marks the degenerate case. 
For typical soft colloidal crystals such as the bcc lattice of charged polystyrene spheres discussed above, one has $C_{11}=1.5$, $C_{12}=0.5$, $C_{44}=0.3$, yielding $\Delta = 0.4 > 0$, which places the system in the $[111]$-dominated regime, consistent with the numerical evaluation.

{
For soft colloidal crystals with relatively small shear moduli $C_{44}$, one usually has $\Delta > 0$, placing the system in the $[111]$-soft regime. 
In contrast, crystals with stronger non-central interactions (e.g., DNA-grafted nanoparticles) can exhibit $C_{12} \approx C_{44}$, leading to $\Delta < 0$ and a $[100]$ softest direction. 
}
The phase diagram therefore provides a straightforward experimental diagnostic: by measuring the three cubic elastic constants, one can immediately predict which crystallographic axis will display the most pronounced static softening as the electrostatic-elastic coupling is enhanced.

{Under the assumption of a fixed particle density, which suppresses homogeneous volume changes, the acoustic softening thresholds determine the first mode to become unstable. The phase diagram in Fig.~\ref{fig:phasediagram} therefore represents the direction of the first acoustic instability under this constraint.}

\subsection{Connection to experiments and unstable strain modes}
The static softening predicted here would manifest as a reduction in the longitudinal sound velocity $v_L(\hat{\mathbf{k}})$. 
In the long-wavelength limit, $v_L^2(\hat{\mathbf{k}}) \propto 1/K(\hat{\mathbf{k}}) - \lambda_g$. 
Thus, as $\lambda_g$ approaches $\lambda_g^{c}(\hat{\mathbf{k}})$, the sound velocity along that direction drops toward zero. 
{Brillouin light scattering or inelastic X-ray scattering can measure directional sound velocities \cite{Cheng2006, Still2008}, and a pronounced anisotropy in the softening could be detected, provided the measurement frequency is low enough that the ionic atmosphere remains in quasi-equilibrium.}

The nature of the instability can be further characterized by examining the eigenvector associated with the vanishing eigenvalue of $\widetilde{M}(\mathbf{k})$. 
For a critical wave vector $\mathbf{k} = k \hat{\mathbf{k}}$, the unstable mode corresponds to a uniform strain $\epsilon_{ij} = \frac{1}{2}(k_i A_j + k_j A_i)$ with $A_i$ the polarization vector. 
Along the high-symmetry directions, the unstable strain patterns associated with the vanishing eigenvalue are as follows:
\begin{itemize}
    \item For $\hat{\mathbf{k}} = [100]$, the soft mode is a pure longitudinal strain $\epsilon_{xx}$, leading to a uniaxial expansion/compression along the cube edge. 
    This tetragonal distortion would break the cubic symmetry to tetragonal.
    \item For $\hat{\mathbf{k}} = [110]$, the soft mode involves a combination of longitudinal strain and shear, resulting in an orthorhombic distortion.
    \item For $\hat{\mathbf{k}} = [111]$, the soft mode is a rhombohedral distortion (pure longitudinal along the body diagonal).
\end{itemize}
Since the critical coupling $\lambda_g^c(\hat{\mathbf{k}})$ varies with direction, the first instability to occur as $\lambda_g$ increases will correspond to whichever of these modes has the lowest threshold. 
For the bcc colloidal crystal parameters discussed above, the $[111]$ rhombohedral mode becomes unstable first, whereas for crystals with $\Delta < 0$ the tetragonal $[100]$ mode would be the precursor. 
These symmetry-breaking patterns could be observed in scattering experiments as precursor fluctuations near the instability threshold.

The identification of these specific unstable strain patterns raises an intriguing experimental possibility: 
the electro-elastic coupling constant $\lambda_g$ is directly tunable via the bulk salt concentration $n_0$ through the Debye screening length $\kappa_0^{-1}$ [see Eq.~\eqref{eq:lambda_DH}]. 
Consequently, sweeping the salt concentration in a colloidal crystal with appropriate elastic anisotropy could drive the system continuously toward the instability threshold. 
As $\lambda_g$ approaches the critical value $\lambda_g^c(\hat{\mathbf{k}})$, the corresponding acoustic branch softens, and the crystal becomes increasingly susceptible to the symmetry-breaking distortion associated with that direction. 
Because electrostatic interactions in these systems are inherently reversible and the underlying lattice remains intact until the mean-field instability point, this suggests the possibility of a reversible, thresholdless martensitic transformation induced by ionic strength. 
Unlike conventional martensitic transitions in atomic crystals, which typically exhibit large hysteresis and require significant undercooling, a soft colloidal crystal operating near the electro-elastic instability could undergo a continuous, diffusionless structural change between cubic and tetragonal, orthorhombic, or rhombohedral variants simply by adjusting the ambient salt concentration.

From the perspective of materials science, this mechanism offers a compelling route toward soft, stimuli-responsive metamaterials. 
By engineering the elastic anisotropy (e.g., via DNA-mediated interactions or particle shape) and the coupling strength $\lambda_g$, one could design colloidal crystals that switch their macroscopic shape or phononic properties in response to minute changes in the chemical environment. 
Such ``electro-elastic shape-memory'' behavior, driven entirely by screened electrostatics, would operate without the need for external mechanical loading or temperature cycling, opening avenues for applications in micro-actuators, tunable acoustic waveguides, and reconfigurable photonic crystals.

\section{Limitations and outlook}

The analysis presented above is based on a continuum description with a local coupling $\lambda_g (\nabla\cdot\mathbf{u})^2$. 
Several extensions and caveats deserve mention.

First, real colloidal crystals are discrete. 
At wave vectors comparable to the Brillouin zone boundary, the local elastic description breaks down and a full lattice-dynamical calculation is required. 
The long-wavelength criterion derived here should be interpreted as a diagnostic for the \emph{onset} of softening, not as a precise prediction of the mode that goes soft at finite $k$.

Second, we have assumed that the electrostatic-elastic coupling is isotropic, i.e., the coupling constant $\lambda_g$ is independent of direction. 
In highly anisotropic crystals, the deformation of the ionic cloud may itself be strongly anisotropic. 
In such cases $\lambda_g$ should be promoted to a second-rank tensor $\lambda_{ij}$, and the coupling term becomes $\lambda_{ij} \theta_i \theta_j$ after suitable coarse-graining. 
The static stability condition would then involve a generalized eigenvalue problem. Exploring the consequences of a tensorial coupling constant remains an interesting direction for future work.

Third, the Gaussian (quadratic) approximation neglects non-linear elastic effects and the possibility of a first-order transition before the quadratic instability is reached. 
Anharmonic terms in the elastic free energy can stabilize the lattice even when the harmonic modulus becomes negative, leading to a weakly first-order transition. 
{ A full nonlinear analysis, including the computation of the free energies of competing crystal structures to determine the thermodynamically preferred phase, is beyond the scope of this work but constitutes a natural extension of the present stability analysis.}

Fourth, the microscopic derivation of $\lambda_g$ in Sec.~\ref{sec:micro_derivation} employed a spherical cell model and the linearized Debye-H\"uckel approximation. 
{
Tamashiro and Schiessel~\cite{Tamashiro2003} have rigorously shown that the linearization of the PB functional in the spherical cell model is asymptotically exact only in the weak-coupling (high-temperature) limit and leads to spurious phase separation and negative compressibilities when applied outside its range of validity. 
Our analytical DH expression, Eq.~\eqref{eq:lambda_DH}, must therefore be applied with caution, and its quantitative predictions are reliable only for systems with moderate charge densities. 
For highly charged colloids where nonlinear screening and charge renormalization become important~\cite{Alexander1984}, a numerical solution of the full nonlinear PB equation is necessary to obtain an accurate $\lambda_g$. 
Within its domain of applicability, however, the DH result provides a transparent, physically motivated link between the phenomenological coupling and the experimental control parameters, and it correctly captures the softening trend.
}

The spherical cell approximation further neglects the faceted shape of the Wigner-Seitz cell. 
For dilute crystals (volume fraction $\lesssim 10^{-3}$) the error is small, but a more accurate calculation using the full PB equation in a periodic geometry would be required for dense systems or for capturing direction-dependent corrections to $\lambda_g$. 
The current derivation assumes a conformally scaling cell, which treats $\lambda_g$ as a scalar. 
Under significant shear strain, the breaking of local $O(3)$ symmetry induces a non-radial polarization of the ionic atmosphere. 
For dense suspensions where the Debye length is comparable to the interparticle spacing, this local field asymmetry necessitates non-diagonal corrections to the effective elastic tensor, leading to a tensor-valued $\lambda_{ijkl}$ and the hybridization of longitudinal and transverse phonon branches. 

{
Finally, we emphasize that the stability analysis in this work focuses on the loss of strong ellipticity (acoustic phonon softening) and is strictly valid under a fixed-volume or fixed-density constraint. 
If the constraint is relaxed, the homogeneous Born instability, which is isotropic, may preempt the directional acoustic instabilities discussed here. 
A complete stability analysis without kinematic constraints remains an open problem.

Furthermore, the numerical examples in Sec.~V use experimentally measured elastic moduli as representative values for the reference constants $C_{ijkl}$. While the qualitative predictions of the phase diagram are independent of the precise origin of the elastic constants, a fully quantitative comparison with experiment would require an independent determination of the reference contribution to the measured moduli, which is beyond the scope of the present theoretical analysis. The DH linearization condition $|y(r)|\ll 1$ has been verified for the representative parameter set ($Z=500$, $\kappa_0 a\approx 2$) through the weak-coupling estimates and the explicit surface potential evaluation presented in Sec.~II B; for other parameter choices, the validity of the DH approximation should be checked on a case-by-case basis.

In this work, we stress that the present theory addresses the local mechanical stability of an already formed crystalline reference state. When an elastic eigenvalue vanishes, the quadratic analysis only signals that the original crystal has lost stability; it does not predict whether the system transforms into another crystal structure, develops an inhomogeneous state, or melts into a liquid. Determining the post-instability fate requires a nonlinear free-energy comparison that lies beyond the scope of the current mean-field stability analysis.} Our analysis is grounded in a mean-field static stability framework. For extremely soft colloidal crystals, thermal fluctuations may renormalize the elastic constants and shift the critical coupling $\lambda_g^c$ from its mean-field value. Near the instability threshold, the softening of the longitudinal acoustic mode leads to a diverging vibrational amplitude $\langle u^2 \rangle$, which implies that the lattice may satisfy the Lindemann criterion and melt before the mechanical instability is reached. This ``fluctuation-induced premature melting'', reminiscent of the Brazovskii effect, may turn the continuous mean-field softening into a weakly first-order transition. A self-consistent treatment of thermal fluctuations within the anisotropic electrostatic-elastic framework remains an open challenge.

\section{Conclusion}

{
We have derived a simple and explicit criterion for the directional acoustic softening of cubic charged colloidal crystals subject to a scalar electrostatic-elastic coupling. 
}
The critical coupling strength $\lambda_g^{c}$ is expressed directly in terms of the three cubic elastic constants $C_{11}$, $C_{12}$, and $C_{44}$ for the $[100]$, $[110]$, and $[111]$ high-symmetry directions. 
A central and somewhat counterintuitive finding is that, under scalar coupling, the face-diagonal $[110]$ direction can never be the unique softest direction. 
The minimum critical coupling is always achieved along either the cube edge $[100]$ (if $\Delta \equiv C_{11} - C_{12} - 2C_{44} < 0$) or the body diagonal $[111]$ (if $\Delta > 0$), while $[110]$ remains intermediate except at the degenerate boundary $\Delta = 0$. 
This exact ordering emerges from the algebraic structure of the inverse Christoffel matrix and the trace-like nature of the electrostatic-elastic coupling term.

We have complemented the continuum analysis with a microscopic derivation of $\lambda_g$ from the PB theory in a spherical Wigner-Seitz cell. 
In the Debye-H\"uckel limit, an analytical expression is obtained that explicitly links the phenomenological parameter to the particle charge $Z$, the lattice spacing $a$, and the bulk Debye screening constant $\kappa_0$. 
Within its range of validity, this expression predicts $\lambda_g > 0$, i.e.\ electrostatic softening, and provides a pathway to estimate the critical coupling for a given experimental system.

Numerical illustrations using elastic moduli reported for a bcc colloidal crystal of charged polystyrene spheres confirm that the body-diagonal $[111]$ direction is the first to soften. 
We have visualized the full angular dependence of $\lambda_g^c$ via spherical color maps and polar projections, and constructed a phase diagram in the plane of reduced elastic constants $(C_{44}/C_{11}, C_{12}/C_{11})$ that compactly summarizes the softest direction for arbitrary cubic moduli. 
The unstable strain patterns associated with each soft direction have been identified: tetragonal for $[100]$, orthorhombic for $[110]$, and rhombohedral for $[111]$, offering observable signatures for scattering experiments.

{
Our analysis is restricted to the long-wavelength, mean-field level and assumes a fixed density, which suppresses the homogeneous Born instability. 
Nevertheless, the simple formulas and phase diagram derived here offer a straightforward diagnostic tool for experimentalists studying anisotropic colloidal crystals under constrained conditions. 
They provide immediate predictions for the most fragile crystallographic direction once the three cubic elastic constants are known, and establish a solid foundation for future extensions incorporating nonlinear elasticity, finite wave-vector instabilities, tensorial electrostatic-elastic couplings, and thermodynamic competition between different crystal structures.
}

\begin{acknowledgments}
H.W.\ gratefully acknowledges inspiring discussions with Rudolf Podgornik during the early stages of this work, which profoundly shaped the theoretical framework presented here. 
This article is dedicated to his memory, in recognition of his pioneering contributions to the statistical physics of Coulomb systems and his generous mentorship to the soft matter community. 
H.W.\ is supported by the General Program of National Natural Science Foundation of China (NSFC) under Grant No.\ 12374210, the open research fund of Songshan Lake Materials Laboratory No.\ 2023SLABFN20, and the startup fund No.\ WIUCASQD2022005 from the Wenzhou Institute University of Chinese Academy of Sciences (WIUCAS). 
Z.C.O.Y.\ is supported by the Major Program of NSFC under Grant No.\ 22193032.
\end{acknowledgments}

\appendix
\renewcommand{\thesection}{\Alph{section}}
\setcounter{section}{0}
\renewcommand{\theequation}{\thesection.\arabic{equation}}
\setcounter{equation}{0}

\section{Appendix A: General expression for \texorpdfstring{$K(\hat{\mathbf{k}})$}{K(k)} in a cubic crystal%
\label{app:generalK}}

For completeness we provide the explicit algebraic form of the directional compliance $K(\hat{\mathbf{k}}) = \hat{\mathbf{k}}^\mathrm{T} \Gamma^{-1}(\hat{\mathbf{k}}) \hat{\mathbf{k}}$ for a cubic crystal with elastic moduli $C_{11}$, $C_{12}$, and $C_{44}$. 
Let $\hat{\mathbf{k}} = (l,m,n)$ with $l^2+m^2+n^2=1$. 
The Christoffel matrix $\Gamma(\hat{\mathbf{k}})$ is given in Eq.~\eqref{eq:Gamma_comp}. 
Its inverse can be obtained by Cramer's rule, and after straightforward algebra one finds the components of $\Gamma^{-1}$ as
\begin{align}
(\Gamma^{-1})_{11} &= \frac{1}{\Delta} \Big[ \Gamma_{22}\Gamma_{33} - \Gamma_{23}^2 \Big], \tag{A.1} \\
(\Gamma^{-1})_{22} &= \frac{1}{\Delta} \Big[ \Gamma_{11}\Gamma_{33} - \Gamma_{13}^2 \Big], \tag{A.2} \\
(\Gamma^{-1})_{33} &= \frac{1}{\Delta} \Big[ \Gamma_{11}\Gamma_{22} - \Gamma_{12}^2 \Big], \tag{A.3} \\
(\Gamma^{-1})_{12} = (\Gamma^{-1})_{21} &= \frac{1}{\Delta} \Big[ \Gamma_{13}\Gamma_{23} - \Gamma_{12}\Gamma_{33} \Big], \tag{A.4} \\
(\Gamma^{-1})_{13} = (\Gamma^{-1})_{31} &= \frac{1}{\Delta} \Big[ \Gamma_{12}\Gamma_{23} - \Gamma_{13}\Gamma_{22} \Big], \tag{A.5} \\
(\Gamma^{-1})_{23} = (\Gamma^{-1})_{32} &= \frac{1}{\Delta} \Big[ \Gamma_{12}\Gamma_{13} - \Gamma_{23}\Gamma_{11} \Big], \tag{A.6}
\label{eq:invGamma}
\end{align}
where $\Delta \equiv \det \Gamma(\hat{\mathbf{k}})$. 
The quantity $K(\hat{\mathbf{k}})$ is then
\begin{align}
K(\hat{\mathbf{k}}) =& l^2 (\Gamma^{-1})_{11} + m^2 (\Gamma^{-1})_{22} + n^2 (\Gamma^{-1})_{33} \tag{A.7} \\
&+ 2lm (\Gamma^{-1})_{12} + 2ln (\Gamma^{-1})_{13} + 2mn (\Gamma^{-1})_{23}. \nonumber
\label{eq:K_general}
\end{align}
For the high-symmetry directions, this reduces to the simple expressions given in Sec.~IV.

\section*{Data Availability Statement}
No data are available for this theoretical research.

\section*{Conflicts of Interest}
The authors declare no conflicts of interest.

\bibliographystyle{apsrev4-2}
\bibliography{aipsamp}

\end{document}